# Equivalence of the Compton and Bethe-Feynman Yield Formulas

Edwin Lyman[a]*

[a]*Union of Concerned Scientists, Washington, DC*

1350 I St NW, Washington, DC, USA

*E-mail: elyman@ucs.org; ORCID ID: http://orcid.org/0000-0001-5388-4143

# Equivalence of the Compton and Bethe-Feynman Yield Formulas

The Bethe-Feynman formula has proven useful for estimating the explosive yields of simple fission weapons. However, derivations of the Bethe-Feynman formula found in the open literature utilize approximations whose validity is not obvious, and which contain an apparent contradiction. Here we show that the Bethe-Feynman formula can also be extracted in a straightforward manner from the results of an earlier approach for estimating yield by Arthur H. Compton, as presented in the 1941 National Academy of Sciences Uranium Committee report. Compton's detailed derivation provides a more intuitive understanding of the physical basis for the Bethe-Feynman formula.

## I. INTRODUCTION

The Bethe-Feynman formula for the yield of a pure fission nuclear weapon, utilizing a core of fissionable material with no tamper/reflector, can be written as [1]

$$Y = KMR^2\alpha^2\delta, \tag{1}$$

where $K$ is a scaling constant, $M$ is the fissionable mass of the core, $R$ is the radius of the core at the time of explosion, $\alpha$ is the peak prompt neutron multiplication rate (where $1/\alpha$ is the "e-folding time"), and $\delta$ is the fractional increase in radius during the post-explosion expansion at which point the core becomes subcritical.

For weapons with a low degree of supercriticality, the efficiency, which is the fraction of the fissionable mass of the core $M$ that is converted to fission energy, can then be written as

$$f = Y/\epsilon M \tag{2}$$

where $\epsilon$ is the energy release per unit of mass fissioned (approximately 17 kt/kg for uranium and 19 kt/kg for plutonium).

In Ref. 1, it is shown that, in certain limits, the 1943 Bethe-Feynman formula is equivalent, within scaling factors, to other formulas for fission weapon yields obtained by Frisch and Peierls in 1940, Serber in 1943, and Pryce and Dirac (year unknown). Ref. 1 also refers to work on the same problem by Compton, which was presented in the declassified 1941 "Report to the President of the National Academy of Sciences by the Academy Committee on Uranium" [2], but does not attempt to show the relationship between the Compton formula and the others. Here we extend the results of Ref. 1 by showing that, with one caveat, the Bethe-Feynman formula can be easily derived from Compton's result.

## II. FISSION WEAPON YIELD AND THE BETHE-FEYNMAN FORMULA

A derivation of the Bethe-Feynman formula can be found in the open literature in the unclassified version of "An Introduction to Nuclear Weapons" by Glasstone and Redman [3]. There it is explained that in a supercritical core (where supercriticality can be achieved through either rapid compression of an implosion device or rapid assembly of subcritical parts in a gun-type device), when a fission chain reaction is initiated by a neutron, the core "incubates" as the energy density increases until it reaches the "explosion time," when the core begins to rapidly expand and the value of $\alpha$ begins to decrease. Fission and significant energy release will continue until after the "second critical," the point where $\alpha$ reaches zero and the core becomes subcritical. The yield is then estimated by approximating the core material as an exponentially expanding ideal gas and calculating the number of e-folding periods that elapse before it becomes subcritical.

The energy release (yield) in kilotons of TNT equivalent that is achieved after $N$ e-folding periods (following initiation of a chain reaction by a single neutron) is given approximately by (Ref. 3):

$$Y \approx (7x10^{-24})\, e^{N}. \tag{3}$$

It makes intuitive sense that the energy per unit mass at the explosion time for a bare (i.e. without a tamper/reflector), uncompressed core should be roughly on the order of that of chemical high explosives, or about 1 kilocalorie/gram, because that is the point when the material begins to behave like a detonating high explosive. For a typical core containing 20 kilograms of fissionable material, this energy density would occur after about 43 e-folding periods. However, in real weapons, the explosion time reportedly occurs well after this point, due to a delay in expansion from the resistance of the tamper and, in implosion devices, the residual inward pressure from the converging shock wave driving the implosion (Ref. 3).

There are varying estimates in the open literature for the actual energy density that is associated with the explosion time. Ref. 3 states that "in general … only a small proportion, usually something like 0.1 percent … of the energy of a fission weapon is produced by explosion time." If we reasonably assume that this estimate for explosion time is associated with a conventional first-generation uranium fission weapon—say, a 20-kt weapon with a 20 kg core—it implies a fission energy release of 20 tons TNT equivalent at explosion time, with a corresponding energy per unit mass ~ 1,000 kcal/g. This energy release would occur at around 50 e-folding periods. To achieve the full yield, another 7 e-folding periods, corresponding to a 1000-fold increase in energy release, would have to elapse after the explosion time.

The weapon's yield is very sensitive to the number of e-folding periods that elapse. A one-period shift results in nearly a factor of three change in yield. Thus an accurate calculation of $t_2$, the time to second critical, is essential for an accurate determination of yield, which partly depends on the number of e-folding periods that occur between the explosion and the second critical. In the derivation of the Bethe-Feynman formula in Ref. 3, $t_2$ is approximated as $1/\alpha$, so that $\alpha t_2 \sim 1$ and thus is a constant. But this is counter-intuitive, or perhaps even paradoxical, because the resulting yield formula would then appear to be independent of the actual number of e-folding periods that occur after the explosion.

## III. THE COMPTON YIELD CALCULATION

In Ref. 2, Arthur H. Compton presented his derivation of the yield of a fission weapon, based on an intuitively plausible physical picture of the explosion [4] similar to the one later described in Ref. 3. In his calculation, Compton estimated the time to second critical $t_2$. But instead of simply taking $t_2$ as inversely proportional to α, he approximated the expansion velocity of the exploding sphere as an exponentially increasing function and expressed the expansion distance to the second critical radius as $\int v(t)\, dt$ over the interval from explosion time to $t_2$.

By relabeling the factor β with the conventional symbol α, and recognizing that the factor 0.12 in Compton's expression is the value of δ for the specific example he presented, Compton's result can be rewritten as

$$t_2 = (1/\overline{\alpha}) \ln(0.18 \delta R^2 \overline{\alpha}^2 \rho_0 / P_0) \tag{4}$$

where the parameters δ and $R$ are the same as those in Eq. (1), $P_0$ is the pressure of the core at explosion time (assuming it has vaporized and behaves like an ideal gas), $\rho_0$ is the initial density at explosion time, and $\overline{\alpha}$ is the time-averaged value of α between explosion time and the second critical, which Compton posits as equal to 0.85α, based on an assumption that the value of $\overline{\alpha}$ is heavily weighted toward its value near the start of the explosion. (To be precise, this was Compton's estimate for the average of the neutron multiplication factor, which is the numerator of α; Compton assumed that the neutron generation time, which is the denominator of α, was a constant over the interval. In fact, the generation time will increase as the core expands and becomes less dense. However, at first approximation accounting for this would affect the value of the 0.85 multiplier but not the assumption that the value is essentially a constant.)

Eq. (4) can be then written as

$$t_2 = {}^{N_2}\!/_{\overline{\alpha}} \tag{5}$$

where $N_2$ is the approximate number of e-folding periods between the explosion time and the second critical. Eqs. (4) and (5) show that $t_2$ indeed depends on the value of $N_2$, which generally will be different for different systems, unlike the assumption in Ref. 3.

## IV. EQUIVALENCE OF THE COMPTON AND BETHE-FEYNMAN FORMULAS

From Eq. (3), the yield in kilotons of TNT equivalent of a fission weapon can be written as

$$Y = (7x10^{-24})\ e^{N_1+N_2} = (7x10^{-24})(e^{N_1})(e^{N_2}) \tag{6}$$

where $N_1$ is the number of e-folding periods between first critical and the explosion time, and $N_2$ is the number between the explosion time and second critical. (Here we are not accounting for the contribution to the final yield from fissions that occur after the second critical, which according to Ref. 3 can be 30% or greater.)

The factor $(7x10^{-2}\ )\ e^{N_1}$ is the total energy release of the core at explosion time (in units of kilotons of TNT equivalent). Using the assumption that the vaporized core at explosion time is an ideal gas, this factor can be substituted for the internal energy of the ideal gas:

$$E_0 = \kappa M P_0/(\gamma - 1)\rho_0 \tag{7}$$

where $\gamma = 5/3$ for a monatomic gas and 4/3 for a photon gas (the actual value of $\gamma$ will depend on the degree of ionization of the gas at explosion time), and $\kappa$ is the appropriate unit conversion factor (e.g. $2.4 \times 10^{-20}$ kilotons TNT/erg for CGS units).

Substituting Eqs. (4), (5), and (7) into Eq. (6), and setting $\overline{\alpha} = \lambda\alpha$, we find that

$$\begin{aligned} Y &= \kappa\left(\frac{MP_0}{(\gamma-1)\rho_0}\right) \times \left(\frac{0.18\delta R^2\overline{\alpha}^2\rho_0}{P_0}\right) \\ &= \kappa\lambda^2\left(\frac{0.18}{\gamma-1}\right)MR^2\alpha^2\delta \end{aligned} \tag{8}$$

which is equivalent to Eq. (1) with a scaling constant $K = 0.18\kappa\lambda^2/(\gamma-1)$, provided that $\lambda$ is itself a constant. This assumption is consistent with Compton's hypothesis that the ratio $\overline{\alpha}/\alpha$ is a fixed value. In actual weapons, $\alpha$ is likely a complicated function of time and it is unlikely that $\overline{\alpha}/\alpha$ would be a universal constant. However, one can show that for a simplified model of an untamped, expanding core, the ratio is a function of the maximum degree of supercriticality, but only varies within a restricted range, and thus it is reasonable to approximate it as a constant.

Interestingly, the pressure at explosion time $P_0$ cancels out of the expression, so that the final yield is independent of the state of the core (including the number of elapsed e-folding periods) at explosion time. This is clearly the result of the oversimplified model and is not likely to be the case for real cores.

With Compton's assumption that $\lambda$=0.85, the scaling constant in Eq. (8) ranges from $0.2\kappa$ to $0.4\kappa$, depending on the value of $\gamma$. These envelop the factor of $1/3(\kappa)$ cited in Ref. 1.

However, a factor of 1/3 underestimates the yield, at least for tamped cores. Compton provided an example of a bare sphere of uranium-235 with a radius of 8.8 cm, corresponding to a mass of 53 kg. Compton believed that this quantity comprised two critical masses, based on his factor-of-two underestimate at the time of the bare critical mass of uranium-235.

Nevertheless, using these parameters as well as Compton's values of 0.4 x$10^8$/sec for $\alpha$ and 0.12 for $\delta$, and a scaling factor of 1/3, Eq. (1) predicts a yield of 6.3 kilotons.

This result can be compared to the known yield for the Hiroshima weapon of around 16 kt [5]. Coincidentally, according to information in the open literature, the uranium-235 content of the Hiroshima core was close to 53 kilograms (at an average enrichment of around 80 percent). However, this quantity corresponded to somewhat more than two critical masses, as the weapon had a thick tungsten carbide tamper/reflector. So it can be seen that for tamped cores, Eq. (1) with a scaling factor of 1/3 underpredicts the actual yield of a Hiroshima-type weapon by a factor of two to three.

Compton intuitively understood that his method would underpredict the yield and adjusted his result with a couple of fudge factors, one of which was a factor of two increase based on an assumption that the fission energy release after second critical would be approximately the same as that before second critical (which according to Ref. 3 may be an overestimate). He also postulated a second factor-of-two increase to account for his having possibly overestimated the expansion pressure at high temperature. Compton also surmised, correctly, that the presence of a tamper could increase the efficiency even further by inhibiting expansion. Perhaps by coincidence, he ended up with an efficiency estimate comparable to that of the Hiroshima bomb, as has been previously pointed out in Ref. 4.[1] However, in hindsight, it is unlikely that the

---

[1] Incidentally, Ref. 4 wrongly claims that Compton made a numerical error in his expression for

yield underprediction could be explained and accurately corrected through these types of simple analyses. An accurate understanding of these effects to properly calibrate the scaling coefficient would require experimental data for validation.

## V. CONCLUSION

The above analysis extends the work of Ref. [1] by showing that, along with other early approaches to estimating the yields of fission weapons, the 1941 Compton approach is also approximately equivalent to the 1943 Bethe-Feynman formula. The Compton approach is also more intuitively clear than other derivations of Bethe-Feynman in the open literature. The Bethe-Feynman formula itself can now be seen as somewhat of a latecomer among the early attempts by various groups to estimate weapon yields.

## DISCLOSURE STATEMENT

The author reports there are no competing interests to declare.

## ACKNOWLEDGMENTS

The author would like to thank Christopher Fichtlscherer of MIT and Prof. Frank von Hippel of Princeton for helpful discussions.

## REFERENCES

[1] J. P. Lestone, M. D. Rosen, and P. Adsley, "Comparison Between Historic Nuclear Explosion Yield Formulas," *Nuclear Technology*, **207,** S1, 352 (2021); https://doi.org/10.1080/00295450.2021.1909372.
[2] "Report to the President of the National Academy of Sciences by the Academy Committee on Uranium," (Nov. 6, 1941); https://www.documentcloud.org/documents/3913458-Report-of-the-Uranium-Committee-Report-to-the.html

---

the kinetic energy of the weapon core in Ref. 2. Compton's expression is consistent with the expression for the average pressure of the core derived in the Appendix of Ref. 3.

[3] S. Glasstone and L.M. Redman. 1972. “An Introduction to Nuclear Weapons.” WASH-1038. US Atomic Energy Commission Division of Military Applications.
[4] B. C. Reed, “Arthur Compton’s 1941 Report on Explosive Fission of U-235: A Look at the Physics,” *Am. J. Phys*., **75**, *12*, 1065 (2007); https://doi.org/10.1119/1.2785193.
[5] Ministry of Foreign Affairs of Japan, “Research Study on Impacts of the Use of Nuclear Weapons in Various Aspects,” March 2014; https://www.mofa.go.jp/files/000051562.pdf (accessed May 30, 2026)